\documentclass[aps,prb,twocolumn,superscriptaddress,showpacs,showkeys,floatfix]{revtex4}

\usepackage{graphicx,color}

\graphicspath{{figs/}}
\begin{document}
\title{Mechanical Strain Effects on Black Phosphorus Nanoresonators}
\author{[Cui-Xia Wang$^{5}$, Chao Zhang$^{5,7}$], Jin-Wu Jiang$^{3\S}$, \\
Harold S. Park$^{4\S}$, T. Rabczuk$^{1,2,5,6\S}$ \\
\emph{\small{$^1$Division of Computational Mechanics, Ton Duc Thang University, Ho Chi Minh City, Vietnam.}}\\
\emph{\small{$^2$Faculty of Civil Engineering, Ton Duc Thang University, Ho Chi Minh City, Vietnam.}}\\
\emph{\small{$^3$Shanghai Institute of Applied Mathematics and Mechanics, Shanghai Key Laboratory of Mechanics in Energy Engineering, Shanghai University, Shanghai 200072, People's Republic of China.}} \\
\emph{\small{$^4$Department of Mechanical Engineering, Boston University, Boston, Massachusetts 02215, USA.}}\\
\emph{\small{$^5$Institute of Structural Mechanics, Bauhaus-University Weimar, 99423 Weimar, Germany.}}\\ 
\emph{\small{$^6$School of Civil, Environmental and Architectural Engineering, Korea University, 136-701 Korea.}} \\
\emph{\small{$^7$College of Water Resources and Architectural Engineering, Northwest A$\&$F University, 712100 Yangling, P.R. China.}} \\
\emph{\small{$^\S$Corresponding author emails: timon.rabczuk@tdt.edu.vn, parkhs@bu.edu, jwjiang5918@hotmail.com}}}
\date{\today}
\begin{abstract}
                                                                                                                                                                                                              
We perform classical molecular dynamics to investigate the effects of mechanical strain on single-layer black phosphorus nanoresonators at different temperatures. We find that the resonant frequency is highly anisotropic in black phosphorus due to its intrinsic puckered configuration, and that the quality factor in the armchair direction is higher than in the zigzag direction at room temperature. The quality factors are also found to be intrinsically larger than graphene and MoS$_{2}$ nanoresonators. The quality factors can be increased by more than a factor of two by applying tensile strain, with uniaxial strain in the armchair direction being most effective.  However, there is an upper bound for the quality factor increase due to nonlinear effects at large strains, after which the quality factor decreases.  The tension induced nonlinear effect is stronger along the zigzag direction, resulting in a smaller maximum strain for quality factor enhancement.

\end{abstract}

\maketitle
\pagebreak

Black phosphorus (BP) is a new two-dimensional nanomaterial that is comprised of atomic layers of phosphorus stacked via van der Waals forces\cite{brown1965refinement}.  BP brings a number of unique properties unavailable in other two-dimensional crystals material. For example, BP has anisotropic properties due to its puckered configuration.\cite{du2010ab,rodin2014strain,low2014tunable,engel2014black}

While most existing experiments have been focused on potential electronic applications of BP\cite{LiL2014,LiuH2014,BuscemaM2014}, a recent experiment showed that the resonant vibration response of BP resonators (BPR) can be achieved at a very high frequency.\cite{wang2015black} However, there have been no theoretical studies on the intrinsic dissipation in BPRs to-date. In particular, it is interesting and important to characterize the effects of mechanical strain on the quality (Q)-factors of BPRs given its anisotropic crystal structure, and furthermore considering that mechanical strain can act as an efficient tool to manipulate various physical properties in the BP structure.\cite{WeiQ2014apl,FeiR2014apl,OngZY2014jppc,SaB2014jpcc,KouL2014ripple,LvHY2014prb,JiangJW2014bpnpr,CaiY2015arxiv} For example, a large uniaxial strain in the direction normal to the SLBP plane can even induce a semiconductor-metal transition.\cite{RodinAS2014prl,HanX2014nl,QinGarxiv14060261,HuangGQ2014arxiv} We thus investigate the mechanical strain effect on the BPRs of armchair and zigzag directions, at different temperatures.

In this work, we examine the effect of mechanical tension on single-layer BPR (SLBPR) via classical molecular dynamical (MD) simulations. Both uniaxial and biaxial tension are found to increase the quality factor of the SLBPR, as the resonant frequency is enhanced by the applied tension. However, the Q-factor decreases beyond a critical strain value due to the introduction of nonlinear energy dissipation, which becomes dominant at large tensile strains.  As a result, there is a critical strain at which the quality factor reaches the maximum value, which is about 4\% and 8\% at 50~K for mechanical tension along the zigzag and armchair directions, respectively. We find that the nonlinear dissipation is stronger if the BPR is stretched along the zigzag direction, which results in a smaller critical strain.

\begin{figure}[tb]
  \begin{center}
    \scalebox{1.0}[1.0]{\includegraphics[width=8cm]{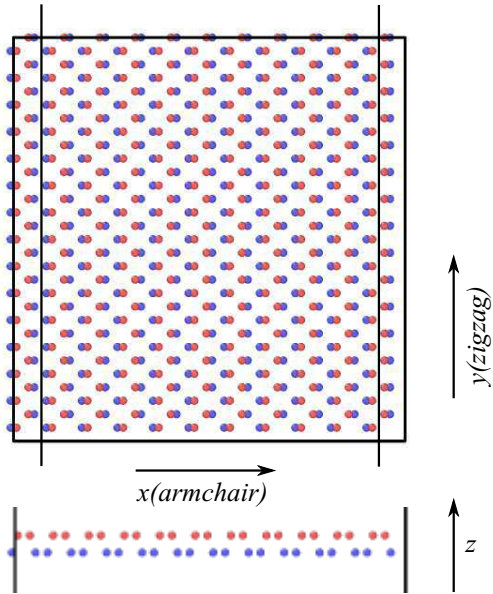}}
  \end{center}
  \caption{(Color online) Configuration of SLBP with dimensions $50\times 50$~{\AA}, from the top view in the top panel, and from the side view in the bottom panel. The total number of atoms is 660.}
  \label{configuration}
\end{figure}

Fig.~\ref{configuration} shows the structure of SLBP of dimension $50\times 50$ ~{\AA} that is used in our simulations. The atomic interactions are described by a recently-developed Stillinger-Weber potential.\cite{jiang2015parametrization} The BPR simulations are performed in the following manner. First, the entire system is thermalized to a constant temperature within the NPT (i.e., the particles number N, the pressure P and the temperature T of the system are constant) ensemble by the Nos\'e-Hoover\cite{Nose,Hoover} thermostat, which is run for 200~{ps}. Second, the SLBP is stretched by uniaxial or biaxial strain along the armchair or zigzag directions. The mechanical strain is applied at a strain rate of $\dot{\epsilon}=0.0001$~{ps$^{-1}$}, which is a typical value in MD simulations. Third, the resonant oscillation of the SLBP is actuated by adding a sine-shaped velocity distribution, $v_0\sin(\pi x_{j}/L)$, to the system. In all simulations, we apply the velocity amplitude $v_0=2.0$~{\AA/ps}, which is small enough to keep the resonant oscillation in the linear region. Fourth, the resonant oscillation of the SLBP is simulated within the NVE (i.e., the particles number N, the volume V and the temperature T of the system are constant) ensemble for 90~ns, and the oscillation energy is recorded to extract the Q-factor.

\begin{figure}[tb]
  \begin{center}
    \scalebox{0.8}[0.8]{\includegraphics[width=8cm]{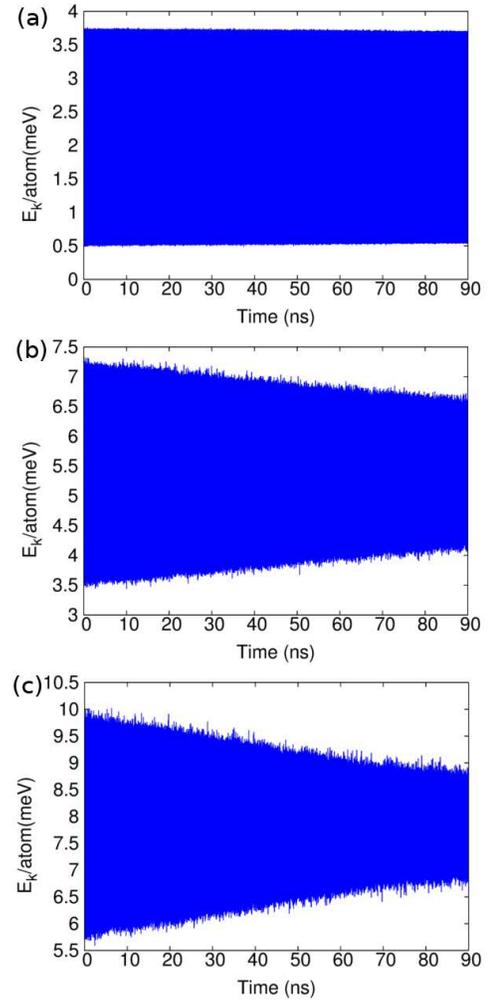}}
  \end{center}
  \caption{(Color online) The kinetic energy per atom for armchair SLBPR at 4.2~K, 30~K and 50~K from top to bottom. The Q-factors are 1621900, 110000 and 63250 respectively.}
  \label{armtem}
\end{figure}

\begin{figure}[tb]
  \begin{center}
  \scalebox{1.0}[1.0]{\includegraphics[width=8cm]{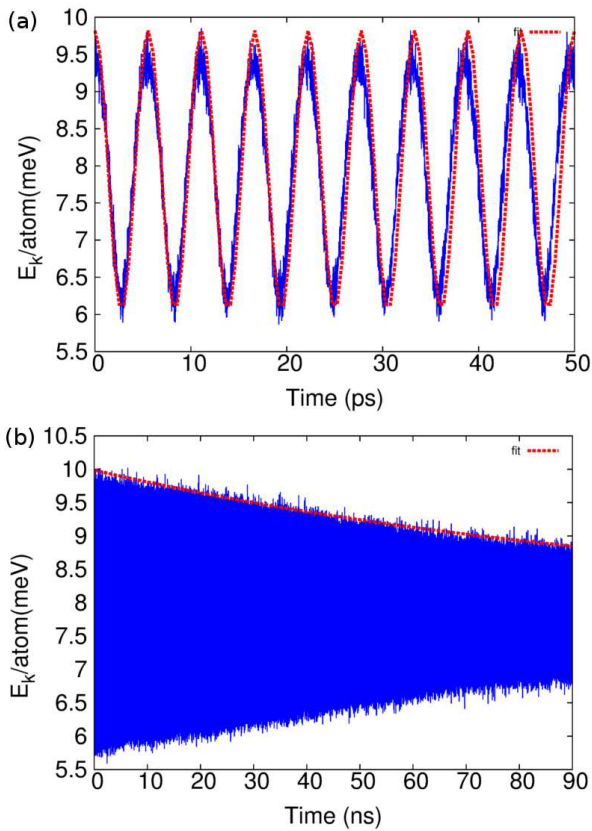}}
  \end{center}
  \caption{(Color online) Two-step fitting procedure to extract the frequency and Q-factor from the kinetic energy time history for armchair SLBPR at 50~K. (a) The kinetic energy is fitted to the function $E_{k}(t)=\bar{E}_{k}+E_{k}^0 \cos(2\pi 2ft)$ in a small time range, giving the frequency $f=0.090874$~THz. (b) The kinetic energy is fitted to the function $E_{k}^{\rm amp}(t)=E_{k}^0 (1-\frac{2\pi}{Q})^{f t}$ in the whole time range, yielding the Q-factor value of 63250.}
  \label{50Kfit}
\end{figure}

\begin{figure}[tb]
  \begin{center}
    \scalebox{1.0}[1.0]{\includegraphics[width=8cm]{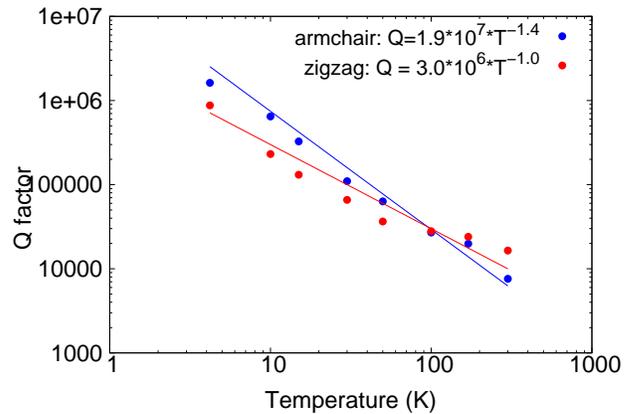}}
  \end{center}
  \caption{(Color online) Temperature dependence for the Q-factors of armchair and zigzag SLBPRs. }
  \label{compareQ}
\end{figure}

\begin{figure}[tb]
  \begin{center}
    \scalebox{1.0}[1.0]{\includegraphics[width=8cm]{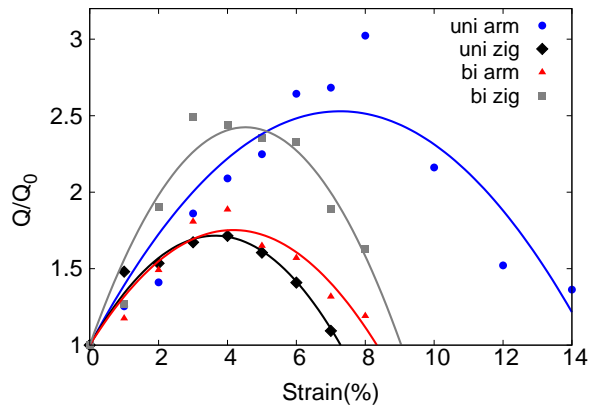}}
  \end{center}
  \caption{(Color online) Strain dependence for the Q-factor of SLBPR in four cases at 50K.  The Q-factor depends on strain as the function $Q/Q_0=-a\epsilon^2+b\epsilon+1.0$, which gives a maximum Q-factor value at a critical strain.}
  \label{fig_qfactor_strain}
\end{figure}

\begin{figure}[tb]
  \begin{center}
    \scalebox{1.0}[1.0]{\includegraphics[width=8cm]{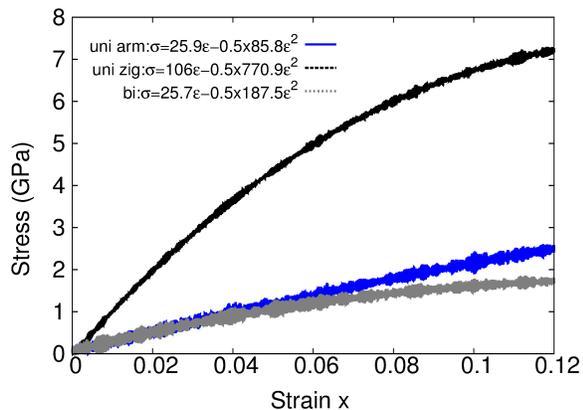}}
  \end{center}
  \caption{(Color online) Stress-strain relation for SLBP under mechanical tension. The stress ($\sigma$) is fitted to a function of strain ($\varepsilon$) as $\sigma=E\varepsilon+\frac{1}{2}D\varepsilon^{2}$, with E as the Young's modulus and D as the TOEC. The nonlinear effect is estimated by the ratio $\gamma=\frac{\frac{1}{2}D}{E}\varepsilon$.}
  \label{stress_strain}
\end{figure}

\begin{figure}[tb]
  \begin{center}
    \scalebox{1.0}[1.0]{\includegraphics[width=8cm]{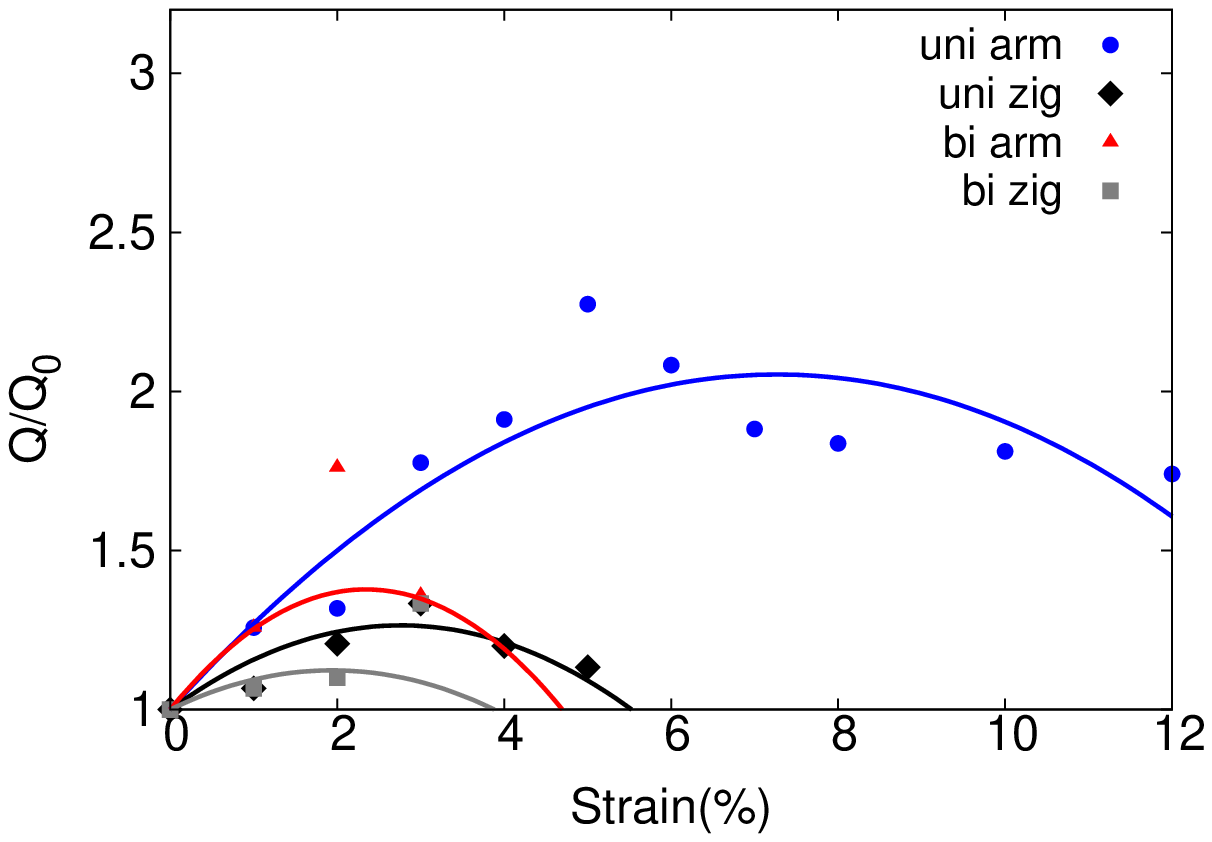}}
  \end{center}
  \caption{(Color online) Strain dependence for the Q-factor of SLBPR in four cases at 170K.}
  \label{170K_qfactor_strain}
\end{figure}

We first examine the intrinsic energy dissipation of the SLBPRs along the armchair and zigzag directions. The intrinsic energy dissipation is induced by thermal vibrations at finite temperatures. Fig.~\ref{armtem} shows the kinetic energy time history in armchair SLBP at 4.2~K, 30~K and 50~K. The oscillation amplitude of the kinetic energy decays gradually, which reflects the energy dissipation during the resonant oscillation of the SLBPR.  As the temperature increases, the energy dissipation becomes stronger, indicating a lower Q-factor at higher temperature.

The frequency and the Q-factor of the resonant oscillation can be extracted from the kinetic energy time history by fitting to the function $E_{k}(t)=\bar{E}_{k}+E_{k}^0 \cos(2\pi 2ft) (1-\frac{2\pi}{Q})^{f t}$. The first term $\bar{E}_{k}$ represents the averaged kinetic energy after the resonant oscillation has completely decayed. The constant $E_{k}^0$ is the total kinetic energy at $t=0$, i.e. at the moment when the resonant oscillation is actuated. The frequency of the resonant oscillation is $f$, so the frequency of the kinetic energy is $2f$. The kinetic energy time history is usually a very long data set, so it is almost impossible to fit it directly to the above function. The fitting procedure is thus done in the following two steps as shown in Fig.~\ref{50Kfit}. First, Fig.~\ref{50Kfit}~(a) shows that the energy time history is fitted to the function $E_{k}(t)=\bar{E}_{k}+E_{k}^0 \cos(2\pi 2ft)$ in a very small time region $t\in [0, 50]$~ps, where the approximation $(1-\frac{2\pi}{Q})^{f t}\approx 1$ has been done for the Q-factor term as the energy dissipation is negligible in the small time range. The parameters $\bar{E}_{k}$, $E_{k}^0$, and $f$ are obtained accurately from this step. Second, Fig.~\ref{50Kfit}~(b) shows that the oscillation amplitude of the kinetic energy can be fitted to the function $E_{k}^{\rm amp}(t)=E_{k}^0 (1-\frac{2\pi}{Q})^{f t}$ in the whole simulation range $t\in [0, 90]$~ns, which determines the Q-factor. Following these fitting procedures, the Q-factor is 63250 for the armchair SLBPR at 50~K.

Fig.~\ref{compareQ} shows the temperature dependence for the Q-factor of the SLBPR along the armchair and zigzag directions. At most temperatures, the Q-factor is larger in the armchair direction. It means that the energy dissipation is weaker for armchair SLBPR, considering that the frequency in the armchair SLBPR is only half of that in the zigzag SLBPR.\cite{jiang2015parametrization} The temperature dependence for the Q-factor can be fitted to the function $Q=1.9\times 10^7 T^{-1.4}$ and $Q=3.0\times 10^6 T^{-1.0}$ for armchair and zigzag SLBPR, respectively. 

These Q-factors are higher than the Q-factors in graphene nanoresonators ($Q=7.8\times 10^4 T^{-1.2}$).\cite{jiang2014mos,kim2009importance}  This is likely because there is also a large energy bandgap in the phonon dispersion of SLBP,\cite{JiangJW2015bpthermal} which helps to preserve the resonant oscillation of the SLBPR.\cite{jiang2014mos}  In contrast, there is no such energy bandgap in the phonon dispersion of graphene, so the SLBPR has higher Q-factor than graphene nanoresonators. The Q-factors of SLBPR's are also higher than those of MoS$_2$ nanoresonators ($Q=5.7\times 10^5 T^{-1.3}$).\cite{jiang2014mos} Both SLBP and MoS$_2$ have energy bandgaps in their phonon dispersions.  This is important as our simulation results imply that nonlinear phonon-phonon scattering is weaker in SLBP, i.e., the resonant energy dissipation is weaker in SLBP than MoS$_2$.

We now report the effects of mechanical strain on both armchair and zigzag SLBPRs at 50~K. We consider four cases, i.e., (I) the effect of uniaxial strain on armchair SLBPR, (II) the effect of uniaxial strain on zigzag SLBPR, (III) the effect of biaxial strain on armchair SLBPR, and (IV) the effect of biaxial strain on zigzag SLBPR. Fig.~\ref{fig_qfactor_strain} shows the strain dependence for the Q-factor (with reference to the value $Q_0$ without strain) of SLBPR under uniaxial or biaxial mechanical tension. In case I, the mechanical strain is applied purely in the armchair direction, while the SLBP is stretched in the zigzag direction in the other three cases.

For all of the four cases, the Q-factor first increases and then decreases after a critical strain value. The Q-factor depends on the strain as the function $Q/Q_0=-a\epsilon^2+b\epsilon+1.0$, where the fitting parameters $(a, b)$ are (0.029, 0.42), (0.055, 0.40), (0.043, 0.36), and (0.070, 0.63) for the four studied cases, respectively. The linear term $b\epsilon$ represents the enhancement effect on the Q-factor by the mechanical tension, as the frequency of the resonator is increased by the tension in the small strain range. The quardratic term $-a\epsilon^2$ is because the Q-factor will be reduced by the nonlinear effect resulting from the mechanical tension in the large strain range. The interplay between these two competing effects results in a maximum value for the Q-factor at a critical strain $\epsilon_c$. The critical strain value is about 8\% for case I, in which the mechanical tension is applied only in the armchair direction. For all other three cases, the critical strain is around 4\%, where the mechanical tension has a component in the zigzag direction.

The differences in the above critical strains can be understood from the different strain induced nonlinear properties in the SLBP. Fig.~\ref{stress_strain} shows the stress-strain curve for the SLBP stretched in the above four cases. The stress-strain curve is fitted to the function $\sigma=E\varepsilon+\frac{1}{2}D\varepsilon^{2}$, with E as the Young's modulus and D as the third-order elastic constant (TOEC)\cite{jiang2015parametrization}. The nonlinear to linear ratio of $\gamma=\frac{\frac{1}{2}D}{E}$ gives an overall estimation of the strain induced nonlinear effect on the SLBP. The parameter $\gamma$ is found to be -1.66 for case I, -3.64 for case II and -3.65 for the other two cases. This means that the nonlinear effect is the weakest in case I, where the SLBP is stretched purely in the armchair direction. As a result, the parameter $a$ has the smallest value for case I, leading to the largest critical strain. This phenomenon (a maximum Q factor due to the strain effect) has also been obtained in nanowire resonators. For example, Kim and Park found that the maximum Q factor occurs around 1.5\% tensile strain in the metal nanowire resonators.\cite{kim2008utilizing}

Fig.~\ref{170K_qfactor_strain} shows the strain effect on the Q-factor at 170~K for all four cases. The critical strain is also observed at this higher temperature, and the critical strain value for SLBR at 170K is about 5\% for case I and around 2-3\% for other three cases. However, the critical strain value is smaller as compared with the critical strain at 50~K in Fig.~\ref{fig_qfactor_strain}.  This is because the nonlinear effect is stronger at higher temperature due to the thermally-induced random vibrations. The combination of the two nonlinear effects (induced by temperature and strain) leads to a smaller critical strain at higher temperature.

In conclusion, we have performed classical molecular dynamics simulations to study the effects of mechanical tension effects on the SLBPR at different temperatures. We find that intrinsically, or neglecting strain, the Q-factors for armchair SLBPR are generally higher than for zigzag SLBPR, and are also larger than those found previously in graphene and MoS$_{2}$ nanoresonators.  When the effects of mechanical strain are considered, our key finding is that there is a maximum point in the strain dependence of the Q-factor due to the competition between the enhancement at small strains and the reduction due to nonlinear effects at large strains.  

\textbf{Acknowledgements} The work is supported by the China Scholarship Council (CXW and CZ). JWJ is supported by the Recruitment Program of Global Youth Experts of China, the National Natural Science Foundation of China (NSFC) under Grant No. 11504225, and the start-up funding from Shanghai University. HSP acknowledges the support of the Mechanical Engineering department at Boston University.

\bibliographystyle{aipnum4-1}
\bibliography{biball}

\end{document}